# Limits to the 1/4 keV Extragalactic X-ray Background

W. Cui[1], W. T. Sanders, and D. McCammon

Department of Physics, University of Wisconsin-Madison, Madison, WI 53706.

S. L. Snowden

USRA/Laboratory for High Energy Astrophysics,

NASA/Goddard Space Flight Center, Greenbelt, MD 20771

AND

D. S. Womble

California Institute of Technology, Astronomy 105-24, Pasadena, CA 91125

## ABSTRACT

We observed several nearby face-on spiral galaxies with the *ROSAT* PSPC. The apparent deficiency in soft X-ray surface brightness observed at the outer portion of their disks is consistent with the absorption of the extragalactic soft X-ray background by material associated with these galaxies, and allows us to place a lower limit on the intensity of this cosmologically important background. From the depth of the soft X-ray shadow observed in NGC 3184, a 95% confidence lower limit was derived to be 32 keV cm$^{-2}$ s$^{-1}$ sr$^{-1}$ keV$^{-1}$ at 1/4 keV. This was obtained by assuming that there is no unresolved 1/4 keV X-ray emission from the outer region of the galaxy which may otherwise partially fill in the shadow: any such emission, or any unresolved structure in the absorbing gas, would imply a larger value. In the deepest exposure to date in this energy range, Hasinger et al. (1993) resolved about 30 keV cm$^{-2}$ s$^{-1}$ sr$^{-1}$ keV$^{-1}$ at 1/4 keV into discrete sources; our current limit is therefore consistent with an extragalactic origin for all of these sources. Our result can also be directly compared with the corresponding upper limit derived from the *ROSAT PSPC* detection of soft X-ray shadows cast by high latitude clouds in Ursa Major, ~65 keV cm$^{-2}$ s$^{-1}$ sr$^{-1}$ keV$^{-1}$ at 1/4 keV. The lower and upper limits are only a factor of 2 apart, and begin to provide a reasonable measurement of the intensity of the 1/4 keV extragalactic X-ray background.

## 1  INTRODUCTION

Despite much attention paid to the extragalactic X-ray background over the past few decades, its origin still remains a mystery. Recently, debate has been focused on whether it originates solely from the integrated emission of known types of active galactic nuclei (AGN), or whether there are also significant contributions from unknown populations of discrete X-ray sources. The X-ray spectrum of this background has been well measured over a large energy range by several experiments (see review articles on this topic by McCammon & Sanders 1990 and Fabian & Barcons 1992, and references therein). Above 60 keV the observed spectrum can be empirically fitted by the sum of two power laws with energy indices of 2.0 and 0.7. Between 3 and 60 keV, the spectrum of thermal bremsstrahlung from a thin intergalactic plasma with kT = 40 keV provides an excellent fit to the data, but this physical origin has been ruled out (Barcons, Fabian, & Rees 1991) by *COBE* measurements (Mather et al. 1990). In the energy range of 3 - 10 keV, the spectrum is well fitted by a power law, $\sim 10 E^{-0.4}$ keV cm$^{-2}$ s$^{-1}$ sr$^{-1}$ keV$^{-1}$, with some 20% uncertainty on the normalization factor derived by different experiments (McCammon & Sanders 1990; Fabian & Barcons 1992; Garmire et al. 1992; Gendreu et al. 1995). At even lower energies, the extragalactic X-ray background is still poorly constrained due to confusion with the bright Galactic soft X-ray emission. A deep *ROSAT* pointing in the Lockman Hole region (Hasinger et al. 1993, H93 hereafter) resolved about 60% of the observed 1-2 keV background into point sources, many of which have been identified as QSOs at redshift of 1-2; all of the background in this energy range can easily be accounted for by extrapolating the measured LogN-LogS relation to a fainter source flux (H93). However, the observed X-ray spectra of these sources (H93), on average, are much softer than that of the diffuse X-ray background), a well-known spectral paradox, so new populations of X-ray sources may still be needed to solve the mystery.

---

[1] Present address: Center for Space Research, Room 37-571, MIT, Cambridge, MA 02139

Limits to the intensity of the 1/4 keV extragalactic X-ray background are important for constraining cosmological models of all types, including the evolution of AGN, which appear to contribute strongly to this background. At 1/4 keV, the large absorption cross section of interstellar matter encourages us to discriminate against the strong foreground emission by measuring soft X-ray shadows cast by high-latitude Galactic clouds or by external galaxies.

The soft X-ray shadows cast by high-latitude clouds are among the earliest *ROSAT* discoveries. They clearly show that a significant fraction of the observed 1/4 keV X-ray background originates a few hundred parsecs away from us, certainly outside of the Local Bubble (Snowden et al. 1990). The derived *distant* 1/4 keV flux varies by a factor of 2 in two fields toward the Draco nebula separated by only about 4° (Snowden et al. 1991; Burrows & Mendenhall 1991), and is a factor of 3-5 higher than that in the direction of Ursa Major (Snowden et al. 1994a). It is the variation in the distant X-ray flux that provides direct evidence for the Galactic halo emission, on top of a homogeneous extragalactic X-ray background. If we take the minimum observed *distant* flux as an upper limit to the isotropic 1/4 keV extragalactic X-ray background, we get an intensity of ~65 keV cm$^{-2}$ s$^{-1}$ sr$^{-1}$ keV$^{-1}$ at 1/4 keV after correction for absorption by the diffuse H II layer in our Galaxy (Reynolds 1991), assuming the equivalent absorbing column density of the H II layer in the Ursa Major field is $\sim 3.4 \times 10^{19}$ H cm$^{-2}$ (Snowden et al. 1994a). One major disadvantage of this approach to studying the extragalactic X-ray background is obviously the uncertainty caused by X-ray emission from the Galactic halo. On the other hand, observing the shadows of external galaxies allows us to study the truly extragalactic soft X-ray background.

This type of experiment was first attempted in the observations of the Small Magellanic Cloud (SMC) (McCammon 1971; McCammon et al. 1976) and M 31 (Margon et al 1974) employing mechanically-collimated proportional counters on sounding rocket experiments. Neither experiment detected any soft X-ray shadows, and the lack of a shadow in the SMC experiment determined a 95% confidence upper limit of ~45 keV cm$^{-2}$ s$^{-1}$ sr$^{-1}$ keV$^{-1}$ on the intensity of the extragalactic X-ray background at 1/4 keV (or ~30 keV cm$^{-2}$s$^{-1}$ sr$^{-1}$ keV$^{-1}$ without correction for absorption by gas associated with the H II layer; see McCammon & Sanders 1990). As shown on the recently published 1/4 keV *ROSAT* all-sky map (Snowden et al. 1995), there is complicated soft X-ray foreground or halo emission in the SMC region that could weaken this result, although even on these much more detailed maps there is no obvious evidence for the anticorrelation of SMC H I and X-ray intensity.

More recently, Barber and Warwick (1994) failed to detect any shadows in *ROSAT* PSPC observations of a tidal plume in NGC 4747 and the extended H I disk in NGC 4725, thus deriving a 95% confidence upper limit of 62 keV cm$^{-2}$ s$^{-1}$ sr$^{-1}$ keV$^{-1}$ at 1/4 keV (or 41 keV cm$^{-2}$ s$^{-1}$ sr$^{-1}$ keV$^{-1}$ without correction for absorption by gas associated with the H II layer) after adding back in the calculated contribution from the resolved extragalactic point sources they removed. Unfortunately, this result is uncertain due to possible filling-in of the shadows by soft X-ray emission from discrete sources in the galaxies below their source detection threshold (~$1.7 \times 10^{-14}$ ergs cm$^{-2}$ s$^{-1}$). These authors included in their analysis the entire NGC 4747, rather than just the associated tidal plume, and the portion of the disk in NGC 4725 that is fairly close to the central region that shows strong X-ray emission. Moreover, they chose to use the H I data from a single-dish measurement in their analysis, instead of the single-dish-calibrated interferometric data of higher spatial resolution (Wevers et al. 1984); the observed clumpier H I distributions in these galaxies might also affect their result.

We observed several nearby face-on spiral galaxies with the *ROSAT* PSPC primarily to study their 0.1 - 2 keV diffuse X-ray emission. The arcminute angular resolution of the *ROSAT* PSPC allowed effective point source removal, thus reducing the unresolved emission from the X-ray sources associated with these galaxies. The analysis and results of that study are presented in the companion paper (Cui et al 1995, Paper 1 hereafter). We also attempted to measure soft X-ray shadows of the extragalactic background cast by these galaxies and the results are presented here. A positive detection of such shadows only allows us to derive a *lower* limit for the soft extragalactic background for two reasons: X-ray emission by the galaxies themselves may reduce the shadows, and there may be H I "holes" in these galaxies on angular scales smaller than the resolution of the *ROSAT* PSPC and the H I surveys that allow the extragalactic background to shine through, thus reducing the depth of the shadows.

A detail description of our data reduction and analysis is presented in Paper 1, as well as 1/4 keV *ROSAT* images, X-ray surface brightness radial profiles, etc., that are relevant to this work. In Section 2, we derive the lower limits on the 1/4 keV extragalactic X-ray background from the corrected radial profiles of the 1/4 keV emission for each of the observed galaxies obtained in Paper 1, and conclude by summarizing and discussing the results in Section 3.

## 2 RESULTS

In Paper 1, we obtained radial profiles of the 1/4 keV surface brightness for each of the five galaxies. The X-ray images were cleaned according to the procedure described in Snowden et al. (1994b): times of the worst short-term enhancements and scattered solar X-rays were eliminated; non-cosmic particle background, long term enhancements and the remainder of the scattered solar X-rays were modeled and subtracted; and vignetting and other instrumental effects were corrected. Contributions from discrete X-ray sources were removed to the faintest practical flux limit, and additional corrections were made to reduce that effective flux limit to the same value in all directions of the field, taking into account the soft X-ray transmission of the target galaxies. The data were binned in concentric rings such that there are always four rings within the largest visually determined extent of each galaxy, and the radial profile was then divided by the foreground Galactic X-ray transmission. The resulting 1/4 keV radial profiles are shown in Fig. 6 of Paper 1.

In this figure, the deficiency in 1/4 keV X-ray surface brightness, apparent at the outer edge of NGC 3184 (ring 4), should be the real shadow of the *unknown* 1/4 keV extragalactic X-ray background cast by this galaxy, since we have already removed all effects of the *known* 1/4 keV extragalactic background by going down to the faintest observed source flux limit using the measured LogN-LogS relation (H93). We now proceed to calculate the surface brightness of this unknown background.

The on-galaxy surface brightness (rings 1-4) in the figure is given by

$$I_{cor}^{i} = I_{gal}^{i} + I_{egu} \times T^{i} + I_{MW}, \quad (1)$$

where $I_{egu}$ is the surface brightness of the *unknown* 1/4 keV extragalactic background, $I_{gal}^{i}$ is the surface brightness of the unresolved 1/4 keV X-ray emission at the i$^{th}$ ring in an observed galaxy, $T^{i}$ is the soft X-ray transmission at that ring, and $I_{MW}$ is the foreground X-ray surface brightness divided by the transmission of the foreground gas in our Galaxy; and the off-galaxy baseline level (rings 8-16) is given by

$$I_{cor}^{base} = I_{egu} + I_{MW}. \quad (2)$$

Subtracting equation (2) from equation (1), we obtain

$$I_{cor}^{i} - I_{cor}^{base} = I_{gal}^{i} - (1 - T^{i})I_{egu}. \quad (3)$$

Solving for, we have

$$I_{egu} = (I^i_{gal} - (I^i_{cor} - I^{base}_{cor}))/(1 - T^i). \qquad (4)$$

Note that

$$I_{cor} = \frac{I_{obs}}{T_f} - I_{rem}, \qquad (5)$$

where $I_{obs}$ is the observed surface brightness, $T_f$ is the foreground X-ray transmission, and $I_{rem}$ is the integrated surface brightness of the remaining *ROSAT* sources.

We know from Paper 1 that the unresolved 1/4 keV emission is bright at the center and fades rapidly toward the outer edges of the observed galaxies. There is little X-ray emission outside of the second ring, so here we simply assume that there is no unresolved 1/4 keV emission at all at ring 4, i.e., $I^4_{gal} = 0$. From equation (4) it is obvious that any such emission would only make $I_{egu}$ larger. Solving equations (4) and (5) simultaneously using the results in Tables 3-5 and Fig. 5 of Paper 1, we derived lower limits to the *unknown* 1/4 keV extragalactic background, shown in Table 1. In order to compare our lower limits with the upper limits obtained from the SMC observations, we added back in the contribution from the *known* extragalactic sources with unabsorbed 0.5 - 2 keV source flux greater than $0.25 \times 10^{-14}$ ergs cm$^{-2}$ s$^{-1}$, which was found to be 30 keV cm$^{-2}$ s$^{-1}$ sr$^{-1}$ keV$^{-1}$ at 1/4 keV in Paper 1. In the table, $I_{egu\_H\,I}$ (or $I_{eg\_H\,I}$) represents the lower limit to the surface brightness of the *unknown* (or *total*) 1/4 keV extragalactic background, derived without correction for soft X-ray absorption by gas associated with the H II layer; and $I_{egu\_H\,II}$ (or $I_{eg\_H\,II}$) is the lower limit derived with the foreground H II layer taken into account.

Snowden & Pietsch (1995) derived a value of $28 \pm 10$ keV cm$^{-2}$ s$^{-1}$ sr$^{-1}$ keV$^{-1}$ for the 1/4 keV extragalactic background, based on the data in one quadrant of the 1/4 keV X-ray image of M101 that contains an apparent "depression" in X-ray surface brightness; this absorption feature appears to be associated with one spiral arm of the galaxy. It is reassuring to see that their result is larger than, but reasonably consistent with what we have derived from the same observation but using the data from all four quadrants in a systematic way, $21 \pm 8$ keV cm$^{-2}$ s$^{-1}$ sr$^{-1}$ keV$^{-1}$ (see Table 1).

Among the five galaxies that we observed, the deepest shadow was found toward NGC 3184. We derive a 95% confidence lower limit on the intensity of the 1/4 keV extragalactic X-ray background by reducing the observed shadow of NGC 3184 by 1.64 times its uncertainty (from Table 1). We presume that the shadows measured toward the other four galaxies are smaller because of more emission from those galaxies, so their shadows are not used in the intensity lower limit determination. The 95% confidence lower limit is 32 keV cm$^{-2}$ s$^{-1}$ sr$^{-1}$ keV$^{-1}$ (26 keV cm$^{-2}$ s$^{-1}$ sr$^{-1}$ keV$^{-1}$ if not correcting for the uncertain absorption by Galactic H II gas). This is a conservative lower limit in that our analysis ignores diffuse X-ray emission from NGC 3184 and possible H I "holes" near its outer edge, both of which would push this value upward.

Table 1: Results[*]

| Galaxy | $I_{egu\_H\,I}$ | $I_{eg\_H\,I}$ | $I_{egu\_H\,II}$ | $I_{eg\_H\,II}$ |
|---|---|---|---|---|
| NGC 3184 | 26 ± 18 | 56 ± 18 | 43 ± 25 | 73 ± 25 |
| M101 | -9 ± 8 | 21 ± 8 | -2 ± 11 | 28 ± 11 |
| NGC 4395 | -3 ± 13 | 27 ± 13 | 6 ± 18 | 36 ± 18 |
| NGC 4736 | 0 ± 23 | 30 ± 23 | 6 ± 31 | 36 ± 31 |
| NGC 5055 | -33 ± 12 | -3 ± 12 | -39 ± 17 | -9 ± 17 |

[*]All intensities are in units of keV cm$^{-2}$ s$^{-1}$ sr$^{-1}$ keV$^{-1}$ at 1/4 keV. The conversion factors used to obtain these values from the units of $10^{-6}$ count s$^{-1}$ arcmin$^{-2}$ are 0.0806 for the R12L band and 0.0723 for the R12 band, which assumes a power law spectrum of $E^{-0.96}$ from H93.

## 3 CONCLUSIONS

A soft X-ray shadow has been detected at the outer edge of NGC 3184 where the emission from this galaxy appears to be very faint. But we failed to detect any shadows in NGC 5055, and also found that the depth of shadows observed in galaxies NGC 4736, M101 and NGC 4395 were all lower than that in NGC 3184. To be consistent, this must be ascribed to a higher X-ray emission level filling in the shadows in these galaxies.

We have derived a 95% confidence lower limit on the intensity of the 1/4 keV extragalactic background to be 32 keV cm$^{-2}$ s$^{-1}$ sr$^{-1}$ keV$^{-1}$ (or 26 keV cm$^{-2}$ s$^{-1}$ sr$^{-1}$ keV$^{-1}$ without correction for absorption by gas associated with the H II layer) from the depth of the shadow observed in NGC 3184, as shown in Table 1. H93 resolved about 30 keV cm$^{-2}$ s$^{-1}$ sr$^{-1}$ keV$^{-1}$ at 1/4 keV into discrete sources; our current limit is consistent with an extragalactic origin for all of these sources, and most likely requires additional contribution from either faint members of the same population below the source detection limit in the *ROSAT* deep survey or other unknown extragalactic sources. This lower limit can also be directly compared with the corresponding upper limit derived from the *ROSAT PSPC* detection of soft X-ray shadows cast by high latitude clouds in Ursa Major, ~65 keV cm$^{-2}$ s$^{-1}$ sr$^{-1}$ keV$^{-1}$, as well as with the less certain limits derived from the SMC (45 keV cm$^{-2}$ s$^{-1}$ sr$^{-1}$ keV$^{-1}$) and from observations of NGC 4725 and NGC 4747 (62 keV cm$^{-2}$ s$^{-1}$ sr$^{-1}$ keV$^{-1}$). The lower and upper limits are only a factor of 2 apart and begin to provide a reasonable measurement of the actual value of the 1/4 keV extragalactic X-ray intensity.

We would like to thank R. Braun for providing the 21 cm data on NGC 4736 in advance of publication and for useful discussions, J. Kamphuis for providing the M101 H I data in digital format and

R. Sancisi for useful discussions. We are grateful to D. P. Cox for many valuable suggestions and discussions on the content and structure of the paper, and to R. Edgar and R. Smith for providing the results of the Raymond-Smith model. We also wish to thank the referee, Daniel Wang, for comments that resulted in an improved manuscript. In determining the long-term enhancement contamination, we partly relied on the *ROSAT* all-sky survey data provided to us courtesy of Max-Planck-Institut für Extraterrestrische Physik. We have also made use of the NASA/IPAC Extragalactic Database and the HEASARC archival database. This work has been supported in part by NASA grants NAG5-629 and NAG5-1665. Support for D. S. W. was provided by NASA through grant number HF-1040.01-92A from the Space Telescope Science Institute, which is operated by the Association of Universities for Research in Astronomy, Inc., under NASA contract NAS5-26555.